\documentclass{IEEEcsmag}

\usepackage[colorlinks,urlcolor=blue,linkcolor=blue,citecolor=blue]{hyperref}
\expandafter\def\expandafter\UrlBreaks\expandafter{\UrlBreaks\do\/\do\*\do\-\do\~\do\'\do\"\do\-}
\usepackage{upmath,color}

\usepackage{cite}
\usepackage{amsmath,amssymb,amsfonts}
\usepackage{algorithmic}
\usepackage{graphicx}
\usepackage{textcomp}
\usepackage{xcolor}
\newcommand{\commentout}[1]{}

\makeatletter
\let\myorg@bibitem\bibitem
\def\bibitem#1#2\par{%
  \@ifundefined{bibitem@#1}{%
    \myorg@bibitem{#1}#2\par
  }{%
    \begingroup
      \color{\csname bibitem@#1\endcsname}%
      \myorg@bibitem{#1}#2\par
    \endgroup
  }%
}

\jvol{XX}
\jnum{XX}
\paper{8}
\jmonth{Month}
\jname{Publication Name}
\jtitle{Publication Title}
\pubyear{2024}

\begin{document}

\sptitle{Theme Article: Special Issue on Certifying and Regulating AI/ML-Centric Applications}

\title{Towards regulatory compliant lifecycle for AI-based medical devices in EU: Industry perspectives} 

\author{Tuomas Granlund}
\affil{Solita and University of Tampere}

\author{Vlad Stirbu}
\affil{CompliancePal and University of Jyväskylä}

\author{Tommi Mikkonen}
\affil{University of Jyväskylä}

\markboth{THEME/FEATURE/DEPARTMENT}{THEME/FEATURE/DEPARTMENT}

\begin{abstract} 
Despite the immense potential of AI-powered medical devices to revolutionize healthcare, concerns regarding their safety in life-critical applications remain. While the European regulatory framework provides a comprehensive approach to medical device software development, it falls short in addressing AI-specific considerations. This article proposes a model to bridge this gap by extending the general idea of AI lifecycle with regulatory activities relevant to AI-enabled medical systems. 

\end{abstract}

\maketitle

\chapteri{M}anufacturing medical devices is a highly regulated industry, and manufacturers must conform to the regulatory requirements of the region where a medical device is being marketed for use. From the European regulatory perspective, the requirements do not distinguish between physical devices or standalone software with a medical purpose, which is, therefore, classified as a medical device in its own right. At the same time, artificial intelligence (AI) and machine learning (ML) based systems permeate nearly every aspect of healthcare, revolutionizing how health is being diagnosed, treated, and managed. However, the unique aspects of AI-based medical systems can also bring new risks or consequences for healthcare professionals and patients. 

To answer the above challenges, the EU Commission is currently formulating an industry-independent regulatory approach and legislation governing AI \cite{ai-act-press}. Parallel to this, the regulatory framework in the medical device sector is also evolving accordingly. Despite the continuous development of the framework, organizations developing AI-enabled medical devices cannot afford to passively wait for regulatory guidance to ensure the safety and effectiveness of their life-critical systems. Instead, a proactive and comprehensive approach is imperative, encompassing the entire medical product lifecycle, from initial research to post-market surveillance and eventual retirement.

This article explores the regulatory implications of AI-enabled medical devices, highlighting the key characteristics and the related challenges for demonstrating conformity and product safety. Understanding these aspects is essential for successfully bringing such devices to market seamlessly. Moreover, to address the pointed-out challenges, the article presents a lifecycle model that integrates established elements of the compliant medical software development model and the crucial lifecycle management aspects for AI systems. The approach is explicitly process-oriented rather than product-oriented to guarantee applicability across various applications.

This paper is structured as follows. First, we provide an overview of the EU regulatory framework governing medical devices. Next, we examine the fundamentals of medical software product development. Building upon this foundation, we delve into the complexities of incorporating AI technology into medical devices, highlighting the specific challenges faced by manufacturers.  We then introduce our AI-specific medical device lifecycle approach, demonstrating how it addresses these challenges. Finally, we conclude by discussing the practical benefits of aligning development and compliance activities, along with the potential for applying our approach to other regulated industries.

\section{THE EU REGULATORY FRAMEWORK}

\begin{figure*}[t]
    \centerline{\includegraphics[width=36pc]{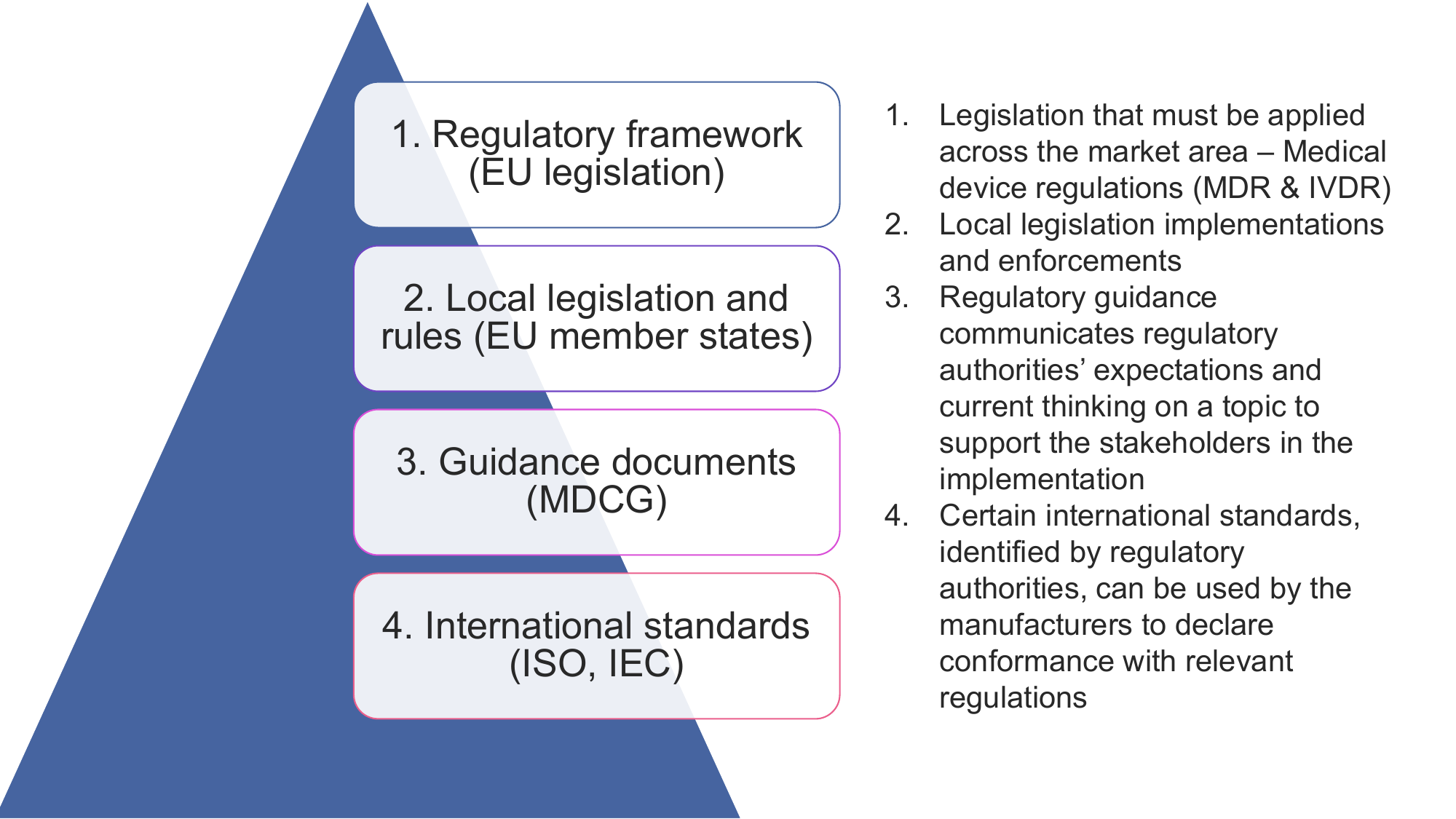}}
    \caption{The regulatory landscape in the medical domain illustrated.}\vspace*{-5pt}
    \label{fig:landscape}
\end{figure*}

The regulatory landscape for product safety in the EU can be interpreted to consist of several layers (Figure~\ref{fig:landscape}). The top layer -- the regulatory framework -- includes legal acts of directives and regulations. There is a significant difference between the two: regulations will enter into force directly in all member states, whereas directives leave a certain amount of leeway to members and may result in variations in different national practices. However, it is also possible to supplement EU regulations with national laws. In the context of medical devices, the applicable EU regulatory requirements are established by Medical Device Regulation (MDR) and In Vitro Diagnostic Medical Device Regulation (IVDR). 

A fundamental characteristic of a large part of Union harmonization legislation is its limited focus on product-specific essential requirements that safeguard the health and safety of users and stakeholders \cite{blue-guide}. The essential product requirements do not impose the precise technical solutions required, specifying only high-level protection outcomes to address the potential hazards related to use and ensure appropriate performance of the product. In the context of medical devices, safety, security, and state-of-the-art clinical performance are the most critical features. 

Specifically related to medical device regulations, the EU Commission provides guidance documents to help manufacturers and other economic operators in meeting the requirements. These guidance documents are adopted by the Medical Device Coordination Group (MDCG), which was created by the Commission to ensure a harmonized implementation of the regulations across the Member States. The MDCG guidance documents are usually developed in active conjunction with the industry stakeholders. While the guidance documents are not legally binding, they provide advice and support for the manufacturers and, at the same time, contribute to the expectation of the regulatory authorities.

In contrast to essential requirements in legislation, technical standards that have been harmonized with the legislation include a certain amount of technical details, taking into account more specific characteristics of a product type. Although using a standard is voluntary, it is highly recommended as it provides a presumption of conformity to those essential requirements that the standard aims to cover. Furthermore, standards are a ready-made and effective tool to address the requirements and are generally expected by the regulatory authorities.

Core standards applicable for all medical software products include the following: (i) general requirements for health software product safety (IEC 82304-1) \cite{iec82304}, (ii) software life cycle process (IEC 62304 \cite{iec62304}), (iii) risk management process (ISO 14971 \cite{iso14971}), and (iv) usability engineering (IEC 62366-1 \cite{iec62366-1}). Furthermore, the manufacturers are expected to have a quality management system that must comply with further associated regulations -- requirements of the Medical Device Quality Systems standard ISO 13485 \cite{iso13485}. These standards aim to cover every phase within the product lifecycle: risk management, design, development, manufacturing, and post-market processes, including maintenance. It is important to note that additional standards may apply based on specific product characteristics. Although the total page count of the list of standards described above is not extensive as such, the requirements included can feel overwhelming as the standards come with rich, sometimes heavy information.

All the layers mentioned above provide a general framework for medical device development. They also leave the responsibility to tailor the exact details of the development process to the manufacturers.

\section{MEDICAL SOFTWARE PRODUCT DEVELOPMENT}

\subsection{Design controls}

\begin{figure*}
    \centerline{\includegraphics[width=36pc]{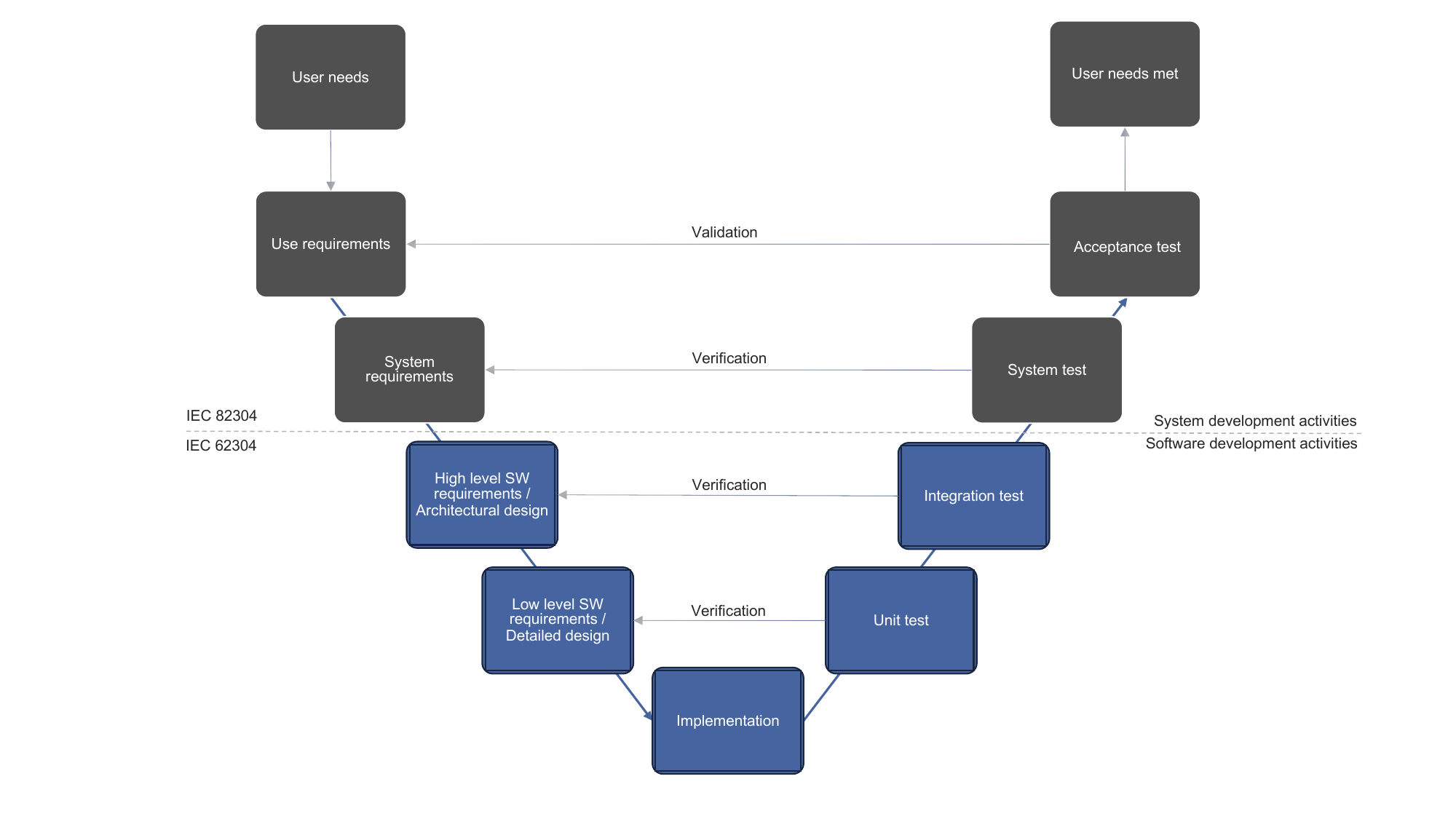}}
    \caption{System and software development design control activities \cite{Stirbu2023}.}\vspace*{-5pt}
    \label{fig:sdlc-activities}
\end{figure*}

As discussed above, the regulatory framework for medical devices aims at ensuring that a medical device is safe to use and clinically effective for its intended medical purpose. To ensure that the goal is achieved, the manufacturers have to follow certain mandatory processes that include control mechanisms for the whole software lifecycle, including the design, development, and manufacturing of the product. In regulatory terms, these processes are referred to as \textit{design controls} and promote a well-designed development process that includes traceability between process inputs and outputs at every stage of the process.

For software medical devices, the design control is implemented in two layers, depicted in Fig. \ref{fig:sdlc-activities}: the system development activities (IEC 82304-1 \cite{iec82304}) and the software development activities (IEC 62304 \cite{iec62304}). At the system level, the identified user needs are converted to system requirements that serve as design inputs for the software development process. During software development, the system requirements are transformed into high-level software requirements that cover the software system and architectural concerns. Later on, the high-level software requirements are further distilled into low-level software requirements that serve as design input for implementation.

The architectural design activity defines the major structural components of the software, known as \textit{software items}. It identifies their key responsibilities, their externally visible properties, and the relationship among them. The resulting software architecture artifact facilitates the correct implementation of the software requirements and is complete when all software requirements can be implemented by the identified software items. The architectural decisions are extremely important for implementing effective risk control measures. The proper understanding and accurate documentation of software items' behavior is essential for ensuring that the software system is safe. Detailed design activities refine the identified software items during architecture design into smaller software items. When a software item is not decomposed further, it is called \textit{software unit}. In the end, the manufacturer is responsible for the granularity of the software decomposition and should ensure that the activity is performed to the appropriate detail to allow a safe and effective implementation.

The resulting code, test cases, and other artifacts, including architecture and detailed module design documentation, are design outputs. Review of the artifacts and the test results act as an effective verification process at the unit, integration, and system levels. Acceptance tests, together with the clinical trial result reports, act as the validation procedure.

\subsection{From plan-driven to continuous engineering}
Although the design control process for medical software is well-defined, the regulatory framework is less opinionated on the actual software development lifecycle that should be used, leaving the freedom to select the best model for the device manufacturers to determine which one suits their products best. This flexibility is essential as the regulatory framework (e.g., IEC 62304) has separate requirements for the development activities performed before the product is placed on the market and the maintenance activities conducted when the product is already on the market. As a result, manufacturers may employ different software development models for these two phases.

Traditionally, manufacturers have adopted plan-driven development methodologies, which were well suited for developing stand-alone hardware-driven products characterized by long lifespans and yearly release lifecycle \cite{long-lifecycle-2013}, intended to be used primarily in hospitals. In general, these products tend to be optimized for a single clinically relevant function, resulting in systems that have lower complexity and reduced dependence and connectivity with third-party systems. Implementing the design control activities in this setup is relatively well-understood, as the relevant standards can be mapped directly to the product and software development stages.

As medical devices became more connected and innovation was delivered at an increasing pace via software, manufacturers started to adopt incremental and iterative software development processes and continuous software engineering practices (e.g., DevOps), which proved successful for mainstream software development. However, implementing the design control activities in an agile process is not straightforward despite attempts to crystallize the experiences of early adopters into formal guidelines, e.g., TIR45 \cite{tir45}. A key challenge is bridging the cultural gap between software development teams, who often embrace agile methodologies, and regulatory professionals, who may be accustomed to more traditional processes. While the software developers are used to working in a highly automated streamlined environment that is optimized for delivering new features to the end users as soon as they have been developed, the regulatory professionals are still accustomed to a document-centered workflow in which changes must be approved by a change control board, often employing their own tools for controlling processes and managing the documents. As such, the interactions between the two disciplines are seen as brittle, the software developers perceiving the regulatory professionals as a \textit{slowdown} factor, while regulatory professionals perceiving the software developers as \textit{undisciplined} \cite{agile-challenges-md}.

\subsection{Regulatory Operations -- RegOps}
To alleviate the challenges mentioned in the previous sub-section, the emerging Regulatory and Operations (RegOps) methodology proposes a set of practices and tools that aim to integrate regulatory compliance into DevOps to shorten the development lifecycle and continuously deliver high-quality software that meets regulatory standards \cite{ahmed}. The goal is to automate and streamline the compliance practices, making them efficient, consistent, and integrated into the overall development and operations workflows. As such, RegOps involves the use of aligned change management practices at code and iteration levels, relying extensively on automation tools, compliance as code, and other practices that ensure that software development and deployment adhere to regulatory requirements. By embedding the needed compliance checks and design controls into the development pipeline, medical device manufacturers can reduce the risk of non-compliance and improve their readiness to deliver software.

\section{REGULATORY CHALLENGES OF MEDICAL AI SYSTEMS}

\subsection{Regulatory landscape for medical AI systems}
While existing medical device regulatory framework provides a foundation, the rapidly evolving landscape for AI-enabled medical device software systems necessitates tailored approaches to ensure their safety and effectiveness. AI-based systems introduce unique challenges and complexities that currently are not addressed explicitly within the regulatory framework~\cite{oravizio2021}. For example, AI systems capable of continuous learning and adaptation can pose challenges for traditional risk assessment methodologies of the regulatory framework. This is because their risk profiles may evolve post-deployment, which can break the traditional design control paradigm of validating changes before deployment. Additionally, the performance and safety of AI models are heavily dependent on the quality and quantity of their training data. Hidden biases and errors in the data, along with model overfitting or underfitting can lead to unpredictable system behavior. 

The lack of precise AI-related regulatory requirements leads to a lack of common understanding within the sector, the increased ambiguity of existing requirements, and unclear expectations of the regulatory authorities, increasing regulatory uncertainty. These factors create potential challenges in achieving compliance and preparing for regulatory submissions. Therefore, the unique aspects of AI-enabled product development require tailored approaches that streamline compliance efforts, ensuring the safety, security, clinical effectiveness, and timely approval of these applications.

\subsection{Other legislation}

A typical characteristic of the legal landscape in the EU is that it is an inherently complicated entity. In addition to the regulatory framework specific to the medical device sector, there is a forthcoming horizontal legislation, the Artificial Intelligence (AI) Act, that will contain AI-specific rules and requirements that will be applied across different sectors. At the time of this writing, EU legislative bodies have just reached a provisional agreement on the AI Act proposal, whereas the exact details of the new regulation are still in the works \cite{ai-act-press}. 

Furthermore, medical devices incorporating AI may be subject to several other directives and regulations. As AI-enabled medical devices are often data-intensive systems handling highly sensitive health data, their design, development, maintenance, and operation processes must also consider data and cybersecurity-related legislation. Depending on the type of AI application, the data laws to consider include the General Data Protection Regulation (GDPR), the Data Governance Act (DGA), and The Data Act. Finally, the cybersecurity laws to consider include the NIS2 Directive (NIS2) and the Cybersecurity Act (CSA) \cite{biasin2023}. 

\subsection{Design control challenges of AI-based systems}
As outlined in the previous section, addressing AI-specific particularities requires tailored design control approaches to ensure regulatory compliance, safety, and effectiveness. Despite the AI technology presenting itself as software artifacts akin to traditional code, it possesses a unique ability to introduce risks that are not apparent at the code level \cite{ml-tech-dept}. The reliance on data for clinical decision-making necessitates specific design control measures that must consider the entire system, not just individual software components. Addressing these challenges requires, for example, a comprehensive understanding of ethical considerations, systematic practices for transparent data collection, and ongoing bias mitigation efforts. These additional measures become apparent at the surface level of AI-based systems, where the footprint of AI-specific software components remains small. Additionally, while general-purpose AI systems might learn and optimize their performance in real time, this ability to evolve fundamentally challenges the established design control expectations for regulated medical devices, which assume a fairly static risk profile.

As an example design control challenge, an AI-based diagnostic tool initially trained on a dataset primarily representing a specific demographic group might exhibit decreased performance or unintended bias when deployed in a more diverse patient population. To address these challenges, questions regarding data-driven design controls, ethical considerations in AI design, and dynamic risk assessment need to be considered throughout the product development lifecycle.

\section{TOWARDS COMPLIANT LIFECYCLE MODEL FOR MEDICAL AI SYSTEMS}

\subsection{Building on state-of-the-art and established processes} %
Due to the lack of explicit requirements or harmonized standards specifically addressing AI-enabled medical devices, manufacturers should proactively adopt a comprehensive approach to ensure product compliance and safety. This approach should encompass all stages of the AI system lifecycle, including design, development, verification, validation, deployment, operation, monitoring, and retirement. To achieve a generally acknowledged state-of-the-art implementation, as required by the regulatory requirements, a suitable strategy is to utilize a widely recognized, sector-independent standard, supplementing its practices with medical device-specific requirements. Such a proactive approach demonstrates the manufacturer's commitment to conformity and safety, ultimately reducing regulatory uncertainty.

As discussed previously, manufacturers have the flexibility to choose a software development model that best aligns with their needs \cite{tir45}. Modern agile practices like RegOps offer significant advantages for product development in terms of efficiency and consistency, making them a natural foundation for AI-related extensions. Utilizing an existing, compliant, and comprehensive development model is beneficial as AI systems naturally integrate both traditional and AI-driven software components \cite{iso_iec5338}, requiring adaptable processes. Additionally, AI components adhere to the same overarching design controls as traditional medical software; therefore, leveraging proven practices within an established model aids in effectively implementing these regulatory requirements.

\begin{figure*}
    \centerline{\includegraphics[width=\textwidth]{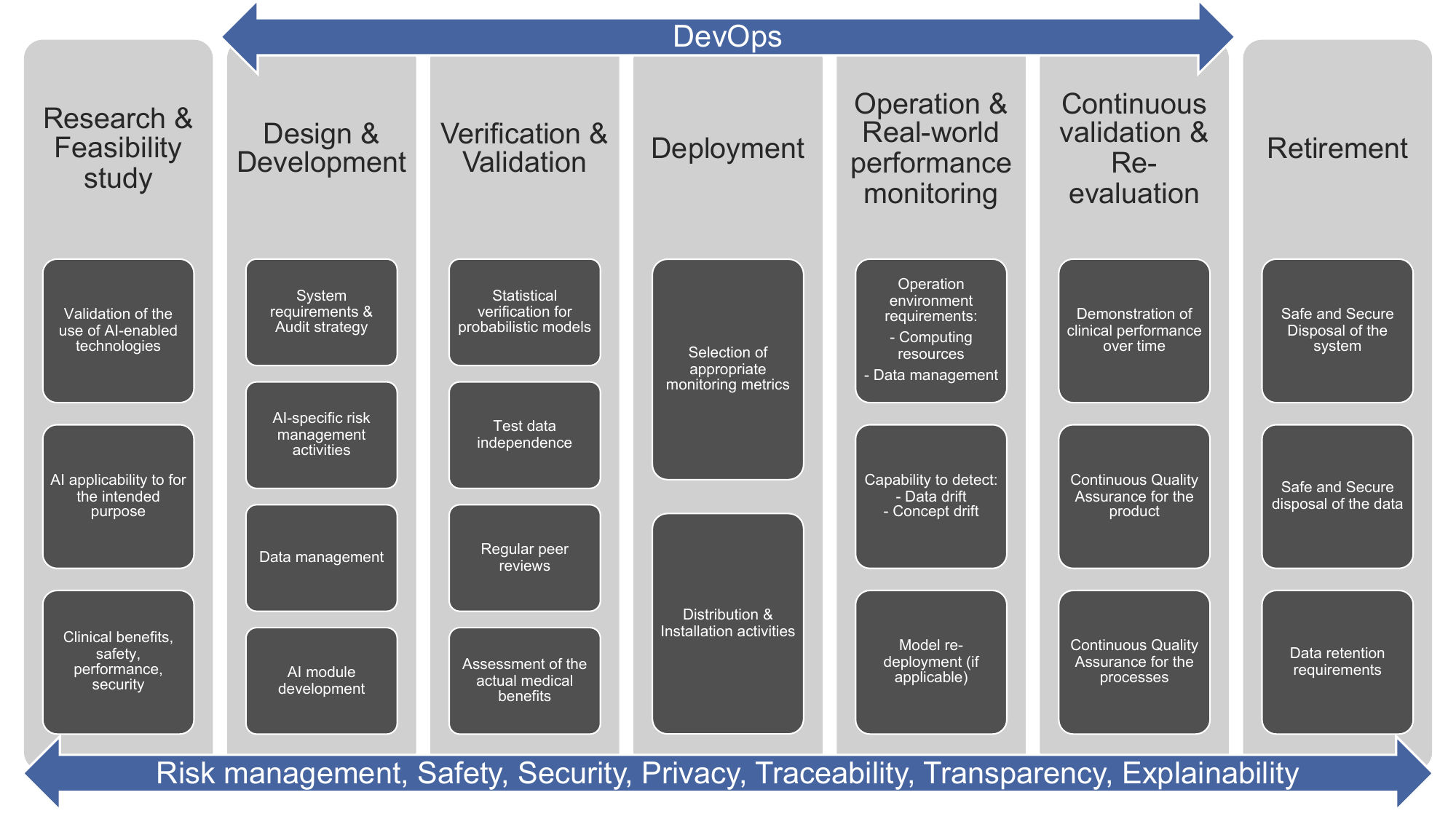}}
    \caption{Lifecycle model for medical device products  including AI technology, adapted from ISO/IEC 5338:2023}\vspace*{-5pt}
    \label{fig:ai-lifecycle}
\end{figure*}

The proposed AI lifecycle model, described in Figure \ref{fig:ai-lifecycle}, was originally developed by Solita in the Agile and Holistic Medical Software Development (AHMED) project \cite{ahmed}. The model is built on top of a RegOps methodology used by Solita for its Oravizio product development~\cite{oravizio2021}, and the lifecycle stages have been adapted from the AI system lifecycle model presented in the ISO/IEC 5338:2023 standard \cite{iso_iec5338}.

\subsection{Lifecycle stages}
As depicted in Figure \ref{fig:ai-lifecycle}, the AI-specific lifecycle model consists of several stages designed to highlight key AI activities from a medical device regulatory perspective and group them into the most meaningful phases. It should be noted that while the stages follow a general chronological order, the activities have cross-stage dependencies and the activities are not completely separated in time. For example, verification activities are often already carried out parallel to development work, and similarly, monitoring capabilities are designed into the system pre-deployment. More generally, activities in later stages build upon the outcomes of earlier ones. Furthermore, the lifecycle allows moving back to earlier stages as needed, and, at any given time, there could be multiple versions of a product, each at its own stage.  Finally, certain activities, such as performance monitoring, are continuous by nature - at least until the AI system is retired. 

Although the model is intended to describe the essential medical device regulatory activities related to AI systems as an of extra layer on top of the general best practices for AI systems, we strongly advise manufacturers to consult and implement all processes and technical practices relevant to their specific product from the ISO/IEC 5338:2023 standard.

\textit{Research and feasibility study} stage aims to validate the use of AI-enabled technology in relation to the intended medical purpose, the clinical risk profile, stakeholder requirements, available resources (e.g. access to high-quality data), and the overall feasibility of the product development project. The intended purpose of the medical device is the single most important aspect to consider at the beginning of the development project as it determines the qualification and classification of the device \cite{pitkanen2020}. In addition, the intended purpose dictates the clinical benefit the system must deliver, which should exceed the associated risks. As AI systems may be more complex than traditional software products, and they may include specific additional risks, the feasibility study should thoroughly assess the potential for AI-enabled technology to be safe, clinically effective, and secure within the scope of the intended medical purpose.

\textit{Design and development} stage aims to capture all AI system requirements and transform them into a functional technical solution. The increased complexity related to the use of advanced technology requires careful consideration of the audit strategy to allow the conformity of the final product to be assessed. One of the most important elements from a compliance perspective is to establish the AI model performance requirements to support the specified intended purpose of the system. The intended purpose and the expected level of clinical performance are closely intertwined with the risks associated with clinical use, and also specific risks unique to AI technology must be addressed under formal risk management procedures. For data-intensive applications, such as models based on machine learning, particular attention must be paid to data management practices as the data can be interpreted as being a raw material for the manufacture of the product. More specifically, the data used in AI module development must be of high quality, traceable to the source, and representative of the intended patient population to ensure the generalisability of results to the population of interest \cite{fda-gmlp-2021}.

\textit{Verification and validation} stage aims to produce objective evidence that the AI module fulfills the specified system and stakeholder requirements, and is able to achieve its intended purpose. In comparison to traditional software systems, which are deterministic by nature, certain AI models may behave in a probabilistic fashion. Therefore, models of this kind require statistical verification techniques to ensure the required correctness and robustness \cite{iso_iec5338}. In accordance to high-quality data management practices, the data sets used for verification must be kept separate from training data \cite{fda-gmlp-2021}. Code reviews are an important method to verify the source code specifically written during the development. However, for certain AI models, such as machine learning models, traditional code reviews may not fully assess functionality. Since algorithms follow steps not directly implemented by humans and depend on dynamic parameters, additional measures are necessary. To mitigate this, regular peer reviews of model architecture, training data, and validation processes should supplement code reviews at appropriate stages throughout development. These reviews offer a fresh perspective, helping uncover hidden errors, ensure robust testing, address potential biases, and promote transparency. By incorporating peer reviews, the verification activities are strengthened, leading to a more reliable and trustworthy AI system. Finally, an assessment needs to be performed in order to prove that the system is able to deliver the intended medical benefits in its intended operational environment.  

\textit{Deployment} stage aims to make the system available for use in its intended operational environment. In the context of medical device software development, there are certain constraints on how and when the product can be delivered and made available for end-use, i.e., clinical use. These constraints are not of a technical nature \cite{oravizio2021} but must be adhered to through established process policy practices. It is essential that exactly the intended software artifact is distributed and installed in a controlled manner and that the installation is verified according to the predefined plan. An important objective of the regulatory framework is to ensure patient safety also after the application has been made available for real-world use. To support this goal, manufacturers are required to systematically and proactively collect and analyze data on the clinical performance of the system \cite{pitkanen2020}. Because AI systems can exhibit performance changes over time, and these changes can vary depending on the specific operational environment, manufacturers need to determine appropriate monitoring metrics \cite{iso_iec5338, fda-gmlp-2021} tailored to each deployment context to effectively track and evaluate these shifts throughout the system's lifecycle.

\textit{Operation and real-world performance monitoring} stage aims to keep the delivered system running and, at the same time, sustain the required performance level. The main focus of monitoring is on the safety and clinical performance of the system \cite{fda-gmlp-2021}. Certain AI systems may require significant computing resources, and in addition, the operation environments may differ from the development environments. Therefore, the requirements related to the operation environment need to be addressed and appropriately monitored. Furthermore, data management practices must address specific requirements related to production data utilized and created by the system, including appropriate production data metrics. For example, the ability to identify unexpected changes in relevant data over time (e.g., data drift) is very important for detecting possible changes in performance. To comprehensively assess the notion of performance degradation, metrics that can effectively identify the influence of extrinsic factors, such as modifications to clinical processes, are essential for promptly detecting potential concept drifts. To counteract any potential performance variations and maintain the intended performance level, the AI system may require relearning or other modifications. When new data is incorporated for retraining, the corresponding test datasets may also necessitate revisions.

\textit{Continuous validation and re-evaluation} stage aims to continuously assess the AI system's clinical performance and ensure it meets the intended level over time. It draws upon technical performance monitoring data from the operation stage and expands the scope to encompass quality assurance processes. This multifaceted approach involves evaluating both the AI system itself and the supporting processes, including development, operation, and maintenance. While technical quality factors are primarily addressed in earlier stages, this stage focuses on aspects related to the quality management system and regulatory compliance. The stage facilitates the creation of comprehensive documentation that serves as objective evidence to support decisions regarding the need for maintenance activities. 

\textit{Retirement} stage aims to oversee the AI system's planned decommissioning, ensuring a controlled and systematic approach to fulfilling all necessary requirements, including informing stakeholders, identifying suitable replacements, archiving relevant data, and responsibly disposing of residual components. Similar to traditional software products, data disposal requires meticulous attention, considering both data security and privacy aspects. However, due to the nature of the data in healthcare systems, data deletion must be carefully coordinated with data retention policies to avoid unintentional compliance violations.
\newline
\newline

\chapteri{A}I systems have the potential to fundamentally change healthcare by enabling advanced capabilities for a wide variety of clinical use cases. However, when used in a safety-critical setting where errors may have serious consequences, it is crucial to ensure that medical applications remain continuously safe and effective in operational use. While the AI gap in the current medical device regulatory framework creates certain uncertainties related to the interpretation of the requirements and expectation level of regulatory authorities, it is of the utmost importance for manufacturers to take proactive approach ensuring product safety throughout the whole lifetime of the product. The lifecycle model for medical device AI systems introduced in this article aims to address this gap. 

The lifecycle model aligns both, product development and compliance activities. Following the stages and activities gives a realistic picture of resources (e.g., personnel and time) that must be allocated in order to build and maintain a compliant medical AI system effectively. The activities require a multi-disciplinary team that performs cross-functional activities. With this complex technology and regulatory requirements, leveraging the established and controlled processes in each discipline is the key to achieving an effective and compliant cooperation model that brings together all development team disciplines into a single track. The lifecycle model follows the RegOps philosophy of shifting the regulatory activities left, by performing them at the time of change. As a result, compliance becomes everyone's concern, and the friction identified between the regulatory and development activities is eliminated.

Although a formal evaluation of the model is beyond the scope of this paper, it's important to note that the proposed model and its activities have been successfully implemented in real-world settings \cite{ahmed}. During the incorporation of AI-based components into a medical device software, the model streamlined the development process, ensuring alignment with regulatory requirements. It provided a clear structure that facilitated, for example, the identification and mitigation of AI-specific risks, enhancing patient safety. Additionally, the model's emphasis on documentation and transparency aided in preparing for regulatory submissions.

While the AI lifecycle model presented in this paper focuses specifically on requirements related to medical products, we believe that it serves as a solid foundation that allows it to applied with other regulatory frameworks. For example, the AI Act developed under the EU's New Legislative Framework approach is consistent with the medical device regulations, suggesting compatibility with our model's structure. We plan to map the requirements of the forthcoming AI Act to this model once the legislative process is complete.  

\def\refname{REFERENCES}
\bibliographystyle{plain}

\begin{thebibliography}{10}

\bibitem{ahmed}
Pasi Ahola, Andrei Baraian, Klaus F{\"o}rger, Tuomas Granlund, Jani Hopia, Risto Kaikkonen, Tommi Mikkonen, Timo Niemirepo, Juha Pajula, Jari Partanen, Timo Pellinen, Vlad Stirbu, and Mika Torhola.
\newblock {\em Agile and Holistic Medical Software Development: Final report of AHMED project}.
\newblock Number VTT-R-01079-22 in VTT Research Report. VTT Technical Research Centre of Finland, Finland, February 2023.

\bibitem{fda-gmlp-2021}
The U.S.~Food \and Drug Administration~(FDA), Health Canada, United~Kingdom’s Medicines, and Healthcare products Regulatory Agency~(MHRA).
\newblock Good machine learning practice for medical device development: Guiding principles, October 2021.

\bibitem{tir45}
{Association for the Advancement of Medical Instrumentation}.
\newblock {AAMI TIR45:2012 (R2018). Guidance On The Use Of AGILE Practices In The Development Of Medical Device Software}, 2018.

\bibitem{biasin2023}
Elisabetta Biasin, Burcu Yaşar, and Erik Kamenjašević.
\newblock New cybersecurity requirements for medical devices in the eu: The forthcoming european health data space, data act, and artificial intelligence act.
\newblock {\em Law, Technology and Humans}, 5(2):43--58, Nov. 2023.

\bibitem{ai-act-press}
{Council of the EU}.
\newblock {Press release: Artificial intelligence act: Council and Parliament strike a deal on the first rules for AI in the world. Accessed 01 May 2024}, 2023.

\bibitem{blue-guide}
{European Commission}.
\newblock {COMMISSION NOTICE - The ‘Blue Guide’ on the implementation of EU product rules 2022}, 2022.

\bibitem{iso_iec5338}
International~Organization for Standardization and International~Electrotechnical Commission.
\newblock { ISO/IEC 5338:2023 - Information technology — Artificial intelligence — AI system life cycle processes}, 2023.

\bibitem{oravizio2021}
Tuomas Granlund, Vlad Stirbu, and Tommi Mikkonen.
\newblock Towards regulatory-compliant {MLOps}: {Oravizio's} journey from a machine learning experiment to a deployed certified medical product.
\newblock {\em SN Computer Science}, 2(5):342, Jun 2021.

\bibitem{iec62304}
{International Electrotechnical Commission}.
\newblock {IEC 62304:2006/A1:2015. Medical device software - Software life-cycle processes}, 2015.

\bibitem{iec62366-1}
{International Electrotechnical Commission}.
\newblock {IEC 62366-1:2015. Medical devices — Part 1: Application of usability engineering to medical devices}, 2015.

\bibitem{iec82304}
{International Electrotechnical Commission}.
\newblock {IEC 82304-1:2016. Health software – Part 1: General requirements for product safety}, 2016.

\bibitem{iso13485}
{International Organization for Standardization}.
\newblock {ISO 13485:2016. Medical devices — Quality management systems — Requirements for regulatory purposes}, 2016.

\bibitem{iso14971}
{International Organization for Standardization}.
\newblock {ISO 14971:2019. Medical devices -- Application of risk management to medical devices}, 2019.

\bibitem{agile-challenges-md}
Martin McHugh, Fergal McCaffery, and Valentine Casey.
\newblock Barriers to adopting agile practices when developing medical device software.
\newblock In Antonia Mas, Antoni Mesquida, Terry Rout, Rory~V. O'Connor, and Alec Dorling, editors, {\em Software Process Improvement and Capability Determination}, pages 141--147, Berlin, Heidelberg, 2012. Springer Berlin Heidelberg.

\bibitem{long-lifecycle-2013}
Canadian~Association of~Radiologists.
\newblock Lifecycle guidance for medical imaging equipment in {Canada}, 2013.

\bibitem{pitkanen2020}
Heikki Pitk\"{a}nen, Leena Raunio, Ilona Santavaara, and Tom St\r{a}hlberg.
\newblock {European Medical Device Regulations MDR \& IVDR - A Guide to Market. Business Finland}, 2020.

\bibitem{ml-tech-dept}
D.~Sculley, Gary Holt, Daniel Golovin, Eugene Davydov, Todd Phillips, Dietmar Ebner, Vinay Chaudhary, Michael Young, Jean-Francois Crespo, and Dan Dennison.
\newblock Hidden technical debt in machine learning systems.
\newblock In {\em Proceedings of the 28th International Conference on Neural Information Processing Systems - Volume 2}, NIPS'15, page 2503–2511, Cambridge, MA, USA, 2015. MIT Press.

\bibitem{Stirbu2023}
Vlad Stirbu, Tuomas Granlund, and Tommi Mikkonen.
\newblock Continuous design control for machine learning in certified medical systems.
\newblock {\em Software Quality Journal}, 31(2):307--333, Jun 2023.

\end{thebibliography}

\vspace*{-8pt}

\begin{IEEEbiography}{Tuomas Granlund}{\,} is a development director and a regulatory compliance specialist at Solita Oy, Finland, and a doctoral student at Tampere University, Finland. His research interests encompass the application of modern software development practices and technologies within regulated domains, with a particular emphasis on streamlining software product development processes for the medical device industry. Contact him at tuomas.granlund@solita.fi. 
\end{IEEEbiography}

\begin{IEEEbiography}{Vlad Stirbu}{\,} is the founder of CompliancePal, Finland, and post doctoral researcher at University of Jyväskylä, Finland. His research interests include software development in regulation-intensive industries and quantum computing. He received his PhD in Software Engineering from Tampere University of Technology, Finland. He is a member of IEEE. Contact him at vlad.stirbu@compliancepal.eu or vlad.a.stirbu@jyu.fi. \vspace*{8pt}
\end{IEEEbiography}

\begin{IEEEbiography}{Tommi Mikkonen} {\,} is a professor of software engineering at the University of Jyväskylä, in Finland. His research interests include continuous software engineering and quantum software. He received his Dr. Tech. in Software Engineerin from Tampere University of Technology, Finland. Contact him at tommi.j.mikkonen@jyu.fi.
\end{IEEEbiography}

\end{document}